\documentclass[twoside]{article}
\catcode`\@=11
\long\def\@makefntext#1{
\protect\noindent \hbox to 3.2pt {\hskip-.9pt
$^{{\eightrm\@thefnmark}}$\hfil}#1\hfill}       

\def\thefootnote{\fnsymbol{footnote}}
\def\@makefnmark{\hbox to 0pt{$^{\@thefnmark}$\hss}}    

\def\ps@myheadings{\let\@mkboth\@gobbletwo
\def\@oddhead{\hbox{}
\rightmark\hfil\eightrm\thepage}
\def\@oddfoot{}\def\@evenhead{\eightrm\thepage\hfil
\leftmark\hbox{}}\def\@evenfoot{}
\def\sectionmark##1{}\def\subsectionmark##1{}}

\oddsidemargin=\evensidemargin
\renewcommand{\thefootnote}{\fnsymbol{footnote}}

\newcounter{sectionc}\newcounter{subsectionc}\newcounter{subsubsectionc}
\renewcommand{\section}[1] {\vspace{12pt}\addtocounter{sectionc}{1}
\setcounter{subsectionc}{0}\setcounter{subsubsectionc}{0}\noindent
    {\tenbf\thesectionc. #1}\par\vspace{5pt}}
\renewcommand{\subsection}[1] {\vspace{12pt}\addtocounter{subsectionc}{1}
    \setcounter{subsubsectionc}{0}\noindent
    {\bf\thesectionc.\thesubsectionc. {\kern1pt \bfit #1}}\par\vspace{5pt}}
\renewcommand{\subsubsection}[1] {\vspace{12pt}\addtocounter{subsubsectionc}{1}
    \noindent{\tenrm\thesectionc.\thesubsectionc.\thesubsubsectionc.
    {\kern1pt \tenit #1}}\par\vspace{5pt}}
\newcommand{\nonumsection}[1] {\vspace{12pt}\noindent{\tenbf #1}
    \par\vspace{5pt}}

\newcounter{appendixc}
\newcounter{subappendixc}[appendixc]
\newcounter{subsubappendixc}[subappendixc]
\renewcommand{\thesubappendixc}{\Alph{appendixc}.\arabic{subappendixc}}
\renewcommand{\thesubsubappendixc}
    {\Alph{appendixc}.\arabic{subappendixc}.\arabic{subsubappendixc}}

\renewcommand{\appendix}[1] {\vspace{12pt}
        \refstepcounter{appendixc}
        \setcounter{figure}{0}
        \setcounter{table}{0}
        \setcounter{lemma}{0}
        \setcounter{theorem}{0}
        \setcounter{corollary}{0}
        \setcounter{definition}{0}
        \setcounter{equation}{0}
        \renewcommand{\thefigure}{\Alph{appendixc}.\arabic{figure}}
        \renewcommand{\thetable}{\Alph{appendixc}.\arabic{table}}
        \renewcommand{\theappendixc}{\Alph{appendixc}}
        \renewcommand{\thelemma}{\Alph{appendixc}.\arabic{lemma}}
        \renewcommand{\thetheorem}{\Alph{appendixc}.\arabic{theorem}}
        \renewcommand{\thedefinition}{\Alph{appendixc}.\arabic{definition}}
        \renewcommand{\thecorollary}{\Alph{appendixc}.\arabic{corollary}}
        \renewcommand{\theequation}{\Alph{appendixc}.\arabic{equation}}
        \noindent{\tenbf Appendix \theappendixc #1}\par\vspace{5pt}}
\newcommand{\subappendix}[1] {\vspace{12pt}
        \refstepcounter{subappendixc}
        \noindent{\bf Appendix \thesubappendixc. {\kern1pt \bfit #1}}
    \par\vspace{5pt}}
\newcommand{\subsubappendix}[1] {\vspace{12pt}
        \refstepcounter{subsubappendixc}
        \noindent{\rm Appendix \thesubsubappendixc. {\kern1pt \tenit #1}}
    \par\vspace{5pt}}

\newcommand{\textlineskip}{\baselineskip=13pt}
\newcommand{\smalllineskip}{\baselineskip=10pt}

\def\eightcirc{
\begin{picture}(0,0)
\put(4.4,1.8){\circle{6.5}}
\end{picture}}
\def\eightcopyright{\eightcirc\kern2.7pt\hbox{\eightrm c}}

\def\abstracts#1#2#3{{
    \centering{\begin{minipage}{4.5in}\footnotesize\baselineskip=10pt
    \parindent=0pt #1\par
    \parindent=15pt #2\par
    \parindent=15pt #3
    \end{minipage}}\par}}

\newcommand{\bibit}{\nineit}

\renewenvironment{thebibliography}[1]
    {\frenchspacing
     \ninerm\baselineskip=11pt
     \begin{list}{\arabic{enumi}.}
    {\usecounter{enumi}\setlength{\parsep}{0pt}
     \setlength{\leftmargin 12.7pt}{\rightmargin 0pt} 
     \setlength{\itemsep}{0pt} \settowidth
    {\labelwidth}{#1.}\sloppy}}{\end{list}}

\newcounter{itemlistc}
\newcounter{romanlistc}
\newcounter{alphlistc}
\newcounter{arabiclistc}

\newcommand{\fcaption}[1]{
        \refstepcounter{figure}
        \setbox\@tempboxa = \hbox{\footnotesize Fig.~\thefigure. #1}
        \ifdim \wd\@tempboxa > 5in
           {\begin{center}
        \parbox{5in}{\footnotesize\smalllineskip Fig.~\thefigure. #1}
            \end{center}}
        \else
             {\begin{center}
             {\footnotesize Fig.~\thefigure. #1}
              \end{center}}
        \fi}

\newcommand{\tcaption}[1]{
        \refstepcounter{table}
        \setbox\@tempboxa = \hbox{\footnotesize Table~\thetable. #1}
        \ifdim \wd\@tempboxa > 5in
           {\begin{center}
        \parbox{5in}{\footnotesize\smalllineskip Table~\thetable. #1}
            \end{center}}
        \else
             {\begin{center}
             {\footnotesize Table~\thetable. #1}
              \end{center}}
        \fi}

\def\@citex[#1]#2{\if@filesw\immediate\write\@auxout
    {\string\citation{#2}}\fi
\def\@citea{}\@cite{\@for\@citeb:=#2\do
    {\@citea\def\@citea{,}\@ifundefined
    {b@\@citeb}{{\bf ?}\@warning
    {Citation `\@citeb' on page \thepage \space undefined}}
    {\csname b@\@citeb\endcsname}}}{#1}}

\newif\if@cghi
\def\cite{\@cghitrue\@ifnextchar [{\@tempswatrue
    \@citex}{\@tempswafalse\@citex[]}}
\def\citelow{\@cghifalse\@ifnextchar [{\@tempswatrue
    \@citex}{\@tempswafalse\@citex[]}}
\def\@cite#1#2{{$\null^{#1}$\if@tempswa\typeout
    {IJCGA warning: optional citation argument
    ignored: `#2'} \fi}}

\def\pmb#1{\setbox0=\hbox{#1}
    \kern-.025em\copy0\kern-\wd0
    \kern.05em\copy0\kern-\wd0
    \kern-.025em\raise.0433em\box0}

\def\fnt#1#2{\footnotetext{\kern-.3em
    {$^{\mbox{\scriptsize #1}}$}{#2}}}

\def\thefootnote{\fnsymbol{footnote}}
\def\@makefnmark{\hbox to 0pt{$^{\@thefnmark}$\hss}}    

\def\ps@myheadings{%
    \let\@oddfoot\@empty\let\@evenfoot\@empty
    \def\@evenhead{\slshape\leftmark\hfil}
    \def\@oddhead{\hfil{\slshape\rightmark}}
    \let\@mkboth\@gobbletwo
    \let\sectionmark\@gobble
    \let\subsectionmark\@gobble
    }
\font\tenrm=cmr10
\font\tenit=cmti10
\font\tenbf=cmbx10
\font\bfit=cmbxti10 at 10pt
\font\ninerm=cmr9
\font\nineit=cmti9

\font\eightrm=cmr8

\def\qed{\hbox{${\vcenter{\vbox{            
   \hrule height 0.4pt\hbox{\vrule width 0.4pt height 6pt
   \kern5pt\vrule width 0.4pt}\hrule height 0.4pt}}}$}}

\renewcommand{\thefootnote}{\fnsymbol{footnote}}  

\begin{document}
\setlength{\textheight}{7.7truein}  

\thispagestyle{empty}


\normalsize\textlineskip

\setcounter{page}{1}



\centerline{\bf CREATION OF THE NONCONFORMAL SCALAR PARTICLES}
\vspace*{0.035truein}
\centerline{\bf IN NONSTATIONARY METRIC}
\vspace*{0.37truein}
\centerline{\footnotesize YU. V. PAVLOV\footnote{E-mail: pavlov@ipme.ru}}
\baselineskip=12pt
\centerline{\footnotesize\it Institute of Mechanical Engineering,
Russian Academy of Sciences,}
\baselineskip=10pt
\centerline{\footnotesize\it 61 Bolshoy, V.O., St.Petersburg, 199178,
Russia}
\vspace*{0.225truein}

\vspace*{0.21truein}
\abstracts{
    The nonconformal scalar field is considered
in $N$-dimensional space-time with metric  which includes,
in particular, the cases of nonhomogeneous spaces and
anisotropic spaces of Bianchi type-I.
   The modified Hamiltonian is constructed.
   Under the diagonalization  of it the energy of quasiparticles is
equal to the oscillator frequency of the wave equation.
   The density of particles created by nonstationary metric is
investigated.
    It is shown  that the densities of conformal and nonconformal
particles created in Friedmann radiative-dominant Universe
coincide.}{}{}

\setcounter{footnote}{0}
\renewcommand{\thefootnote}{\alph{footnote}}

\vspace*{1pt}\textlineskip
\section{Introduction}
\vspace*{-0.5pt}
\noindent
   Quantum field theory in curved space-time was actively developed
in the 70-ties of the last century (see books\cite{1,2}).
   However, there are a number of unsolved problems in this theory.
   One of them is related to the definition of an elementary particle
and vacuum in a curved space-time.
   The reason is that in the case of a curved space  there is no group of
symmetries similar to the Poincar\'{e} group in the Minkowski space.
   If we believe that a particle is associated with a quantum of energy,
then according to quantum mechanics, the observation of particles at
an instant implies finding an eigenstate of the Hamiltonian.
    The Hamiltonian diagonalization method\cite{1} takes this into account
automatically.

    The use of the Hamiltonian constructed from the metrical
energy-momen\-tum tensor, first proposed by A.A.Grib and
S.G.Mamayev,\cite{3} was successful in the homogeneous isotropic
space-time for the conformal scalar fields.
   But such Hamiltonian leads to the difficulties related to an
infinite density of created quasiparticles in the nonconformal
case.\cite{4}
    The energy of such quasiparticles differs from the oscillator
frequency of the wave equation.\cite{5,6}
    The nonconformal case is important for investigation by many reasons.
    In particular,  the additional
nonconformal terms are dominant in the vacuum expectation values
of the energy-momentum tensor.\cite{7}
    Recently,\cite{8} the modified Hamiltonian was found  that the
density of the particles corresponding to its diagonal form and created
in the nonstationary homogeneous isotropic space-time is finite.

    The aim of this paper is the generalization of the method,
proposed in Ref.~8, for the cases of the nongomogeneous and anisotropic
spaces.
    In Sec.~2 we quantize the nonconformal scalar field
in curved space and construct the modified Hamiltonian.
    In Sec.~3 we diagonalize this Hamiltonian and investigate the
questions about the energies and the density of created particles.

\section{Quantizing a Scalar Field in Curved Space}
\noindent
    We consider a complex scalar field  $\varphi(x)$ of the mass $m$
with Lagrangian
    \begin{equation}
L(x)=\sqrt{|g|}\ [\,g^{ik}\partial_i\varphi^*\partial_k\varphi -
(m^2+f(R))\, \varphi^* \varphi \,] \,,
\label{Lag}
\end{equation}
  and corresponding equation of motion
\begin{equation}
 ({\nabla}^i {\nabla}_{\! i} + f(R) + m^2) \varphi(x)=0 \, ,
\label{Eqm}
\end{equation}
   where ${\nabla}_{\! i}$ are the covariant derivatives in
$N$-dimensional space-time,
$ g\!=\!{\rm det}(g_{ik})$,  $\ f(R)$ is the function of the
invariant combinations $g_{ik}$ and curvature tensor $R^i_{\,klm}$.
    The equation~(\ref{Eqm}) is conformally invariant if $m=0$
and $f(R)=\xi_c R $, where $R$ is scalar curvature, and
$\xi_c = (N-2)/\,[\,4\,(N-1)] $ (conformal coupling).
   We investigate the case of metric which in some coordinate system
$(x)=(\eta, {\bf x})$ takes the form
    \begin{equation}
ds^2 = a^2(\eta) \left( d\eta^2 - \gamma_{\alpha \beta}(x)\,
d x^\alpha d x^\beta \right) \,,
\label{gikn}
\end{equation}
   where  $\alpha, \beta= 1, \ldots, N-1 $ and
$\ {\rm det}(\gamma_{\alpha \beta})=\gamma({\bf x})$.
    It realizes, in particular, for the homogeneous isotropic space-time;
\ for the nonhomogeneous spaces with
   \begin{equation}
ds^2=dt^2 - a^2(t) \, \gamma_{\alpha \beta}({\bf x}) \,
dx^\alpha d x^\beta  \ ;
\label{giknh}
\end{equation}
        for the anisotropic metric of Bianchi type-I
$
\left( ds^2 = dt^2 - a^2_\alpha(t)\,(d x^\alpha )^2 \right). \
$
     The equation~(\ref{Eqm}) for the functions
$ \tilde{\varphi}=a^{(N-2)/2}\varphi $
in coordinates $(\eta, {\bf x})$ takes the form
   \begin{equation}
\tilde{\varphi}''- \Delta_{N-1} \tilde{\varphi} +
\biggl[ \left(m^2 + f(R) \right) a^2 - \frac{N-2}{4} \left( 2c'+
(N-2) c^2 \right) \biggr] \, \tilde{\varphi}=0 \,,
\label{eft}
\end{equation}
    where the prime denotes the derivative with respect to time $\eta$,
$\ c \equiv a'/a \ $,
$\ \Delta_{N-1} \equiv \gamma^{-1/2} \partial_\alpha (\sqrt{\gamma}\,
\gamma^{\alpha \beta} \partial_{\beta} ) $.
    Let us assume that Laplace-Beltrami operator $\Delta_{N-1}$
has the complete orthonormal set of the eigenfunctions
$\Phi_{\!J}({\bf x})$ non-depending on time
    \begin{equation}
\Delta_{N-1} \Phi_{\!J}({\bf x}) = -\lambda^2(J,\eta)\,
\Phi_{\!J}({\bf x}) \,.
\label{DelPhi}
\end{equation}
    The full set of the motion equation solutions can be found in the form
$ \tilde{\varphi}(x) = g_J(\eta) \Phi_{\!J} ({\bf x}) $\,,
  where
      \begin{equation}
g_J''(\eta) + \Omega^2(\eta)\,g_J(\eta)=0 \,,
\label{gdd}
\end{equation}
$\Omega$ is the oscillator frequency:
\begin{equation}
\Omega^2(\eta)=\left( m^2 + f(R) \right) a^2 -
\frac{N-2}{4} \left( 2c'+(N-2) c^2 \right) +\lambda^2(J,\eta) \,.
\label{Ome}
\end{equation}
    For quantization, we expand the field $ \tilde{\varphi}(x) $
       \begin{equation}
\tilde{\varphi}(x)=\int \! d\mu(J)\,\biggl[ \tilde{\varphi}{}^{(+)}_J \,
a^{(+)}_J + \tilde{\varphi}{}^{(-)}_J \, a^{(-)}_J \,\biggr] \ ,
\label{fff}
\end{equation}
     where $d\mu(J)$ is the measure on the space of $\Delta_{N-1}$
eigenvalues,
\begin{equation}
\tilde{\varphi}{}^{(+)}_J (x) =\frac{1}{\sqrt{2}}\,
g_J(\eta)\,\Phi^*_{\!J}({\bf x}) \ ,
\ \  \ \
\tilde{\varphi}{}^{(-)}_J (x)=\biggl(\tilde{\varphi}{}_J^{(+)}({\bf x})
 \biggr)^* \,,
\label{fpm}
\end{equation}
     and impose the usual commutation relations on the operators
$ a_J^{(\pm)} , \ \stackrel{*}{a}\!{\!}_J^{(\pm)}$.

    We construct Hamiltonian as canonical one for variables
$\tilde{\varphi}(x)$ and $\tilde{\varphi}^*(x)$.
  If we add $N$-divergence $(\partial J^i/ \partial x^i)$ to Lagrangian
density~(\ref{Lag}), where in the coordinate system $(\eta, {\bf x})$
the $ N$-vector
$\ (J^i)=(\sqrt{\gamma}\,c\,\tilde{\varphi}^*\, \tilde{\varphi}\,(N-2)/2,
\, 0, \, \ldots \,, 0) $, \
   the motion equation~(\ref{Eqm}) is invariant under this addition.
    By using the Lagrangian density
$ L^{\Delta}(x)=L(x)+({\partial J^i}/{\partial x^i})$, \
we obtain by integration on hypersurface $\Sigma :\ \eta=const $
of Hamiltonian density
$ h(x)= \tilde{\varphi}'\,(\partial L^{\Delta})/
(\partial \tilde{\varphi}') +
\tilde{\varphi}^{* \prime}\,(\partial L^{\Delta})/
(\partial \tilde{\varphi}^{* \prime})-L^{\Delta}(x) $,
the modified Hamiltonian
     \begin{eqnarray}
H(\eta) &=& \int_\Sigma h(x) \, d^{N-1}x = \int_\Sigma d^{N-1}x \,
  \sqrt{\gamma} \, \Biggl\{
\tilde{\varphi}^{* \prime} \tilde{\varphi}{}'
+\gamma^{\alpha \beta} \partial_\alpha\tilde{\varphi}^*
\partial_\beta \tilde{\varphi}+
\nonumber                \\
&+& \biggl[ \left(m^2+f(R) \right) a^2 - \frac{N-2}{4} \left(2c'+
(N-2)c^2\right)
\biggr]\,  \tilde{\varphi}^* \tilde{\varphi} \Biggr\} \,.
\label{Hp}
\end{eqnarray}
      Hamiltonian~(\ref{Hp}) is expressed in terms of the operators
$ a_J^{(\pm)} , \ \stackrel{*}{a}\!{\!}_J^{(\pm)}$ by
    \begin{equation}
H(\eta)= \! \int \! \! d\mu(J)  \biggl\{ \! E_J(\eta)\!
\left(\stackrel{*}{a}\!{\!}^{(+)}_J a^{(-)}_J +
\stackrel{*}{a}\!{\!}^{(-)}_{\bar{J}} a^{(+)}_{\bar{J}} \right) +
F_J(\eta)  \stackrel{*}{a}\!{\!}^{(+)}_J a^{(+)}_{\bar{J}} +
F^*_J(\eta) \stackrel{*}{a}\!{\!}^{(-)}_{\bar{J}} a^{(-)}_J \! \biggr\},
\label{H}
\end{equation}
    where
\begin{equation}
E_J(\eta)=\frac{1}{2} \biggl[\, |g_J'|^2+ \Omega^2 |g_J|^2 \,
\biggr]  \ ,
\ \ \ F_J(\eta)=\frac{\vartheta_{\!J}}{2} \biggl[\, g_J'{}^{\! 2} +
\Omega^2  g_J^{\, 2} \, \biggr]  \,,
\label{EJFJ}
\end{equation}
   here we used the normalization condition
$ g_J\,g_J^*{}' - g_J^{\, \prime}\,g_J^* = -2 i \  $ and the
special choice of the eigenfunctions at which for an arbitrary $J$ such
$\bar{J}$ existed, that $\Phi_{\!J}^*({\bf x}) = \vartheta_{\!J}
\Phi_{\!\bar{J}}({\bf x})$.

\section{Hamiltonian Diagonalization and Particle Creation}
\noindent
The Hamiltonian diagonalization for an arbitrary instant $\eta$
is realized in terms of the operators
$ \stackrel{*}{b}\!{\!}^{(\pm)}_J, \  b^{(\pm)}_J $
  related to the operators
$ \stackrel{*}{a}\!{\!}^{(\pm)}_J, \  a^{(\pm)}_J $
   via the time-dependent Bogolubov transformations
\begin{equation}
a_J^{(-)}=\alpha^*_J(\eta) b^{(-)}_J(\eta) -
\beta_J(\eta) \vartheta_{\!J} b^{(+)}_{\bar{J}}(\eta), \ \ \
\stackrel{*}{a}\!{\!}_J^{(-)}=
\alpha^*_J(\eta) \! \stackrel{*}{b}\!{\!}^{(-)}_J\!(\eta) -
\beta_J(\eta) \vartheta_{\!J} \!
\stackrel{*}{b}\!{\!}^{(+)}_{\bar{J}}\!(\eta),
\label{db}
\end{equation}
     where the functions $\alpha_J, \ \beta_J $ satisfy
the initial conditions
$|\alpha_J(\eta_0)|=1, \ \beta_J(\eta_0)=0 $  and the identity
$ |\alpha_J(\eta)|^2-|\beta_J(\eta)|^2=1 $.
   Substituting expansion~(\ref{db}) in  (\ref{H})
and requiring that the coefficients at the nondiagonal terms
$ \stackrel{*}{b}\!{\!}^{(\pm)}_J  b^{(\pm)}_J $
vanish, we obtain
    \begin{equation}
-2\alpha_J\beta_J \vartheta_{\!J}E_J + \alpha_J^2 F_J
+\beta_J^2 \vartheta_{\!J}^2 F_J^* = 0   \,, \ \ \
|\beta_J|^2=
\frac{1}{4\Omega}\biggl( |g_J'|^2+\Omega^2 |g_J|^2 \biggr) -\frac{1}{2} \,,
\label{bFJ}
\end{equation}
     \begin{equation}
H(\eta) =\int d\mu(J) \,\Omega(\eta) \,
\left(\,\stackrel{*}{b}\!{\!}^{(+)}_J b^{(-)}_J +
\stackrel{*}{b}\!{\!}^{(-)}_{\bar{J}} b^{(+)}_{\bar{J}}\, \right) \,.
\label{Hbb}
\end{equation}
   Therefore, the energies of quasiparticles corresponding to the diagonal
form of the Hamiltonian~(\ref{H}) are equal to the oscillator frequency
$\Omega(\eta)$.

   The vacuum state for instant $\eta$ is defined by
$
b^{(-)}_J|\,0_\eta\!> \,=\, \stackrel{*}{b}\!{\!}^{(-)}_J|\,0_\eta\!>\,
= 0 \,.
$
    The state $|\,0\!>=|\,0_{\eta_0}\!>$ contains
$|\beta_J(\eta)|^2$ quasiparticle pairs corresponding to the operators
$b^{(\pm)}_J(\eta)$ in every mode.\cite{1,2}
    The density of created particles is proportional to
$\int d\mu(J)\, |\beta_J|^2 $.
    Using the first iteration in the integral equation for
$S(\eta)=|\beta_J(\eta)|^2$
    \begin{equation}
S(\eta)=\frac{1}{2}\,\int_{\eta_0}^\eta d\eta_1 \,
w(\eta_1)\, \int_{\eta_0}^{\eta_1} d\eta_2 \,w(\eta_2)\,
(1+2 S(\eta_2)) \cos[2\,\Theta(\eta_2,\eta_1)]  \,,
\label{iuS}
\end{equation}
    where $w(\eta)=\Omega'(\eta)/\,\Omega(\eta) \ $,
$\Theta(\eta_1, \eta_2)=\int_{\eta_1}^{\eta_2}
\Omega(\eta)\,d \eta$,
     we can see, that
if the eigenvalues $-\lambda^2(J)$ of the Laplace-Beltrami operator
    $\Delta_{N-1}$ don't depend on time, then
    $w \sim \lambda^{-2}$\ and \
    $S \sim \lambda^{-6} $ in  $\lambda(J) \to \infty $.

    Therefore, in the space-time with metric~(\ref{giknh}) the density
of created particles for $N=4$ is finite. (In Bianchi type-I the
result is infinite).
   From equation~(\ref{iuS}) for $S(\eta)$ we can see
that  the densities of conformal and nonconformal
(under $ f(R)=\xi R$)\ particles  created in Friedmann
radiative-dominant  Universe coincide.

\nonumsection{Acknowledgements}
\noindent
The author is grateful to Prof. A.A.Grib for helpful discussions.
This work was supported by Min. of Education of Russia, grant E00-3-163.

\nonumsection{References}


\begin{thebibliography}{000}
\bibitem{1}
A. A. Grib, S. G. Mamayev and V. M. Mostepanenko, {\bibit
Vacuum Quantum Effects in Strong Fields}
(Friedmann Laboratory Publishing, St.Petersburg, 1994).

\bibitem{2}
N. D. Birrell and P. C. W. Davies, {\bibit Quantum Fields in Curved Space}
(Cambridge University Press, Cambridge, 1982).

\bibitem{3}
A. A. Grib and S. G. Mamayev, {\bibit Yad. Fiz.}
{\bf 10}, (1969) 1276 (Engl. trans. in {\bibit Sov. J. Nucl. Phys.}
{\bf 10}, 722).

\bibitem{4}
S. A. Fulling, {\bibit Gen. Relat. Gravit.} {\bf 10}, (1979) 807.

\bibitem{5}
M. Castagnino and R. Ferraro, {\bibit Phys. Rev.} {\bf D34}, (1986) 497.

\bibitem{6}
V. B. Bezerra, V. M. Mostepanenko and C. Romero,
{\bibit Int. J. Mod. Phys.} {\bf D7}, (1998) 249.

\bibitem{7}
M. Bordag, J. Lindig, V. M. Mostepanenko and Yu. V. Pavlov,
{\bibit Int. J. Mod. Phys.} {\bf D6}, (1997) 449.

\bibitem{8}
Yu. V. Pavlov, {\bibit Theor. Math. Phys.} {\bf 126}, (2001) 92.
\end{thebibliography}
\end{document}